\documentclass[11pt]{article}

\usepackage{lineno,hyperref}

\usepackage{url}  
\usepackage{ifthen}  
\usepackage{multicol}   
\usepackage[utf8]{inputenc} 
\usepackage{amssymb,amsfonts,amsmath}
\usepackage{graphicx}
\usepackage{float}
\usepackage{flafter}
\floatstyle{boxed}
\newcommand{\bdsl}{\boldsymbol}
\newcommand{\ind}{\stackrel{ind}{\sim}}
\newcommand{\iid}{\stackrel{iid}{\sim}}

\usepackage[figuresright]{rotating}
\usepackage[dvips]{color}

\begin{document}

\title{\bf Comparison of Clustering Methods for \\Time Course Genomic Data:\\
Applications to Aging Effects}

\author{ {\sc Yafeng Zhang$^1$, Steve Horvath$^{1,2}$,}\\ 
{\sc Roel Ophoff$^2$ and Donatello Telesca$^1$}\\[.1in]
$^1$UCLA, Biostatistics\\ 
$^2$UCLA, Human Genetics}

\maketitle

\centerline{Abstract}
\noindent Time course microarray data provide insight about dynamic biological processes. While several clustering methods have been proposed for the analysis of  these data structures, comparison and selection of appropriate clustering methods are seldom discussed.
We compared $3$  probabilistic based clustering methods and $3$ distance based clustering methods for time course microarray data.
Among probabilistic  methods, we considered:  smoothing spline clustering also known as model based functional data analysis (MFDA), functional clustering models for sparsely sampled data (FCM) and model-based clustering (MCLUST). Among distance based methods, we considered: weighted gene co-expression network analysis (WGCNA), clustering with dynamic time warping distance (DTW) and clustering with autocorrelation based distance (ACF).
We studied these algorithms in both simulated settings and case study data. Our investigations showed that FCM performed very well when gene curves were short and sparse. DTW and WGCNA performed well when gene curves were medium or long ($>=10$ observations). SSC performed very well when there were clusters of gene curves similar to one another. Overall, ACF performed poorly in these applications. In terms of computation time, FCM, SSC and DTW were considerably slower than MCLUST and WGCNA. WGCNA outperformed MCLUST by generating more accurate and biological meaningful clustering results.
WGCNA and MCLUST are the best methods among the 6 methods compared, when performance and computation time are both taken into account. WGCNA outperforms MCLUST, but MCLUST provides model based inference and uncertainty measure of clustering results.


\section{Background}
Genome wide microarray data record expression levels of thousands genes simultaneously, by measuring the amount of cRNA  attached to DNA sequence probes on a microarray chip. It has been shown that co-regulated genes are also co-expressed \cite{SpellmanSherlock1998, Cho1998}. Therefore, clustering methods identifying groups of co-expressed genes have proven to be powerful tools in the study of genetic functions and biological processes. Similarly, DNA methylation microarray analysis measures methylation levels for thousands of DNA sequence probes. A gene is methylated when a methyl group is attached to its DNA sequence, which, in turn, stops the gene's expression and its biological function. Clustering analysis on DNA methylation data provides us an opportunity to learn what genes are turned off collectively during biological processes. Time course microarray data, where gene expression or methylation levels are measured over time, provide a good opportunity to understand the dynamics of biological systems and how  gene activity evolves during biological processes. Genes with similar profiles are more likely to be co-regulated or involved in the same biological functions. To make full use of time course microarray data, one needs to take into account the time sequence and time dependence of observations.

An important and widely used class of clustering methods is distance based clustering.  These techniques compute distance measures between data points and identify clusters according to partition algorithms. Here, we describe several classic examples. In hierarchical clustering, a matrix storing Euclidean distances between data points is used as the input to the clustering procedure, where close data points form clusters first, and then close clusters form bigger clusters until there is only one cluster. The hierarchical clustering structure can be represented by a dendrogram, a tree diagram in which leaves correspond to data points and branches represent clusters \cite{EisenEtal1998, SpellmanSherlock1998}. In k-means clustering \cite{MacQueen1967, Tavazoie1999}, cluster centroid are provided as input to the procedure, then the following two steps are repeated iteratively, until clustering structure becomes stable: 1. each data point is assigned to a cluster whose centroid it is closest to in terms of certain distance measure; 2. cluster centroids are computed as the mean of data points in each cluster. Weighted Gene Co-expression Network Analysis (WGCNA) uses a biologically more relevant distance measure in hierarchical clustering to form clusters of genes \cite{ZhangHorvath2005, LangfelderHorvath2008}. Specific distance measures can also be defined for functional data and used in distance based clustering methods. For example, Time warping distances are defined by computing distances between time transformation functions that are estimated to model curves' transformed time scales in a functional data registration model \cite{TelescaEtal2009}.

An alternative approach to data partitioning is model-based or probabilistic clustering.  These methods assume that data are sampled from a mixture of underlying probability distributions \cite{FraleyRaftery2002a}. Given the number of clusters, parameter estimation is usually carried out by EM algorithm. The number of clusters can be determined as a model selection problem via some criteria, such as Bayesian information criterion (BIC). Its application to clustering gene expression data was first discussed in \cite{Yeung2001}, where a Gaussian mixture model was used.

Several probabilistic clustering methods have been proposed specifically for functional data, and some of them also find successful applications to time course microarray data \cite{MaEtal2006, MaZhong2008, JamesSugar2003, LuanLi2003, BarJoseph2002, ZhouWakefield2006, GaffneySmyth2003}. In those methods, functional data is commonly modeled as stochastic processes centered on cluster specific shape functions with curve-specific random variations. Different spline regression models are used to represent cluster specific and curve specific shape functions, for example, natural splines \cite{JamesSugar2003}, B-splines \cite{LuanLi2003, BarJoseph2002} and smoothing splines \cite{MaEtal2006, MaZhong2008}. Usually, parameter estimation is carried out via the EM algorithm, and cluster numbers are determined through model selection using some criteria, such as BIC.

While the literature has seen increased interest in the development of several alternative clustering methods, little attention has been paid to a thorough comparison of these techniques, when applied to realistic data scenarios.  In this chapter, we review, in detail, smoothing spline clustering also known as model based functional data analysis (MFDA) \cite{MaEtal2006, MaZhong2008}, functional clustering model for sparsely sampled data (FCM) \cite{JamesSugar2003} and  model-based clustering (MCLUST) \cite{FraleyRaftery2002a}. For distance based clustering, we review Weighted Gene Co-expression Network Analysis (WGCNA) \cite{ZhangHorvath2005}, dynamic time warping distance (DTW) \cite{TelescaEtal2009} and autocorrelation based distance (ACF) \cite{DUrso2009}. We also compare their performances in simulation studies and applications to real datasets. In our analysis we consider methods that have been implemented in \verb'R' or that are easily reproduced.

We extend the applications of these algorithms to cross section studies, where an ordering covariate, specifically age, is concurrently measured. Several of these studies are constrained by the impossibility of repeated access to tissue for molecular interrogation. For example, brain tissue is only available after death. We show how it is potentially useful to define pseudo-longitudinal representations of cross sectional designs. We evaluate two longitudinal representation strategies, the binning method and smoothing method in the context of our comparative studies.

The rest of this paper is organized as follows: In $\S$ \ref{sec:methods} we provide a review of  probabilistic  clustering and distance based clustering techniques and describe the binning method and smoothing method for data transformation. In $\S$ \ref{sec:simulation} we compare different clustering methods with two simulation studies. In $\S$ \ref{sec:casestudy}, we conduct two studies, relating DNA methylation clusters to Polycomb group protein targets (PCGTs) and clustering preservation study between expression datasets, to compare the methods' ability to identify aging-related clusters. In $\S$ \ref{sec:discussion}, we discuss some issues in choosing the methods. Finally, we summarize our findings in $\S$ \ref{sec:conclusion}.

\section{Methods}\label{clustMethod}

To fix notation, we consider gene expression data monitored over a uniform time grid ${\bdsl t} = (t_1, \ldots, t_j, \ldots, t_m)'$, where $m$ is the total number of sampling points.
For each gene we observe a time course expression  vector ${\bdsl y}_i = \Big(y_{i1}, \ldots, y_{ij}, \ldots, y_{im}\Big)'$, $(i=1,\ldots, p)$, where $p$ is the number of genes.

\subsection{Probabilistic Clustering }

\noindent \underline{Smoothing spline clustering}:\\
In smoothing spline clustering, also known as model based functional data analysis (MFDA), gene expression curves are modeled  to include random intercepts within a finite location mixture\cite{MaEtal2006, MaZhong2008}. Specifically, for gene  $i$ at time $t_{j}$ , given cluster membership $k$,  $(k=1,\ldots, K)$, where $K$ is number of clusters, it is assumed that $y_{ij}=\mu_k(t_{j})+b_i+\epsilon_{ij},$
where $\mu_k()$ is the $k$th cluster specific shape function of time, $b_i \ind N(0, \sigma^2_{bk})$  is the gene specific random intercept and  $\epsilon_{ij} \iid \mathrm{N}(0, \sigma^2)$ is the error term.  With $b_i$ integrated out,  the sampling model for  $\bdsl{y}_i$ can be written as a finite mixture
\begin{equation}{\label{MFDAmixModel}}
\bdsl{y}_i \sim \sum_{k=1}^K p_k \,\mathrm{N}_m (\bdsl{\mu}_k, \Sigma_k);
\end{equation}
where $p_k$ is the probability that gene $i$ belongs to cluster $k$, $\Sigma_k=\sigma_{bk}^2\bdsl{1}_m\bdsl{1}_m^\prime+\sigma^2\bdsl{I}_m$
and $\bdsl{\mu_k} = (\mu_k(t_1),\ldots,\mu_k(t_m))'$.
Estimates of $\bdsl{\mu_k}$ are obtained maximizing a penalized likelihood function and the number of clusters, $K$,
is determined via BIC comparison.  This procedure is implemented in the  \verb'R' package \textit{Model Based Functional Data Analysis} \cite{MaZhong2008}.

\vskip.2in
\noindent \underline{Functional clustering model for sparsely sampled data}:\\
In a functional clustering model (FCM)\cite{JamesSugar2003}, gene curve $i$, given cluster membership $k$, is modeled as  $\bdsl{y}_i=S(\bdsl{t})(\bdsl{\lambda_0}+\Lambda \bdsl{\alpha}_k+\bdsl{b}_i)+\bdsl{\epsilon}_i, $ , where $S(\bdsl{t}): m \times q$ is the spline basis matrix,  $\bdsl{\lambda_0}$ is the $s$-dimensional basis coefficient vector for the overall shape function, $\bdsl{\alpha}_k$ is the $h$-dimensional basis coefficient vector for the $k$th cluster specific shape function, and $\Lambda: q\times h$ is the transition matrix to reduce the parameter dimension from $q$ to $h$, where $h \leq \mathrm{min}(q, K-1)$.  The model admits random gene specific coefficients  $\bdsl{b}_i \sim \mathrm{N}_s(\bdsl{0}, D)$ and errors $\bdsl{\epsilon}_i \sim \mathrm{N}_m(\bdsl{0}, \sigma^2\bdsl{I}_m)$. With $\bdsl{b}_i$ integrated out, the marginal sampling model for $\bdsl{y}_i$ is
\begin{equation}{\label{FCMmixModel}}
\bdsl{y}_i \sim \sum_{k=1}^K p_k\,\mathrm{N}_m\Big(S(\bdsl{t})(\bdsl{\lambda_0}+\Lambda \bdsl{\alpha}_k),\Sigma_i\Big)
\end{equation}
where $p_k$ is the probability that gene $i$ belongs to cluster $k$ and $\Sigma_i=\sigma^2\bdsl{I}+S_i D S'_i$.

Under several identifiability conditions,
the model may be fit via EM and the number of clusters $K$ can be
determined through model selection using BIC.  This method is implemented in the \verb'R' package  \textit{Functional Clustering Model}.

\vskip.2in
\noindent \underline{Model-based clustering}:\\
This approach is based on finite location-scale mixtures  \cite{FraleyRaftery2002a,  Yeung2001}, specifically,
\begin{equation}{\label{mbcModel}}
\bdsl{y}_i \sim \sum_{k=1}^K p_k\,\mathrm{N}_m(\bdsl{\mu}_k,\Sigma_k).
\end{equation}
 Maximum likelihood estimation is carried out via EM and the number of clusters $K$ is selected via  BIC.   This method is implemented in the \verb'R' package MCLUST
\cite{FraleyRaftery2002b}.

\subsection{Distance Based Clustering}\label{distanceClust}
\noindent \underline{Weighted Gene Co-expression Network Analysis}: \\
In WGCNA \cite{ZhangHorvath2005}, a topological overlap measure is define between gene expression data vectors and hierarchical clustering is used to cluster genes. There is evidence that the topological overlap measure is superior when it comes to co-expression clustering and networks \cite{RavaszEtal2002}. Specifically, the topological overlap measure between gene $i$ and gene $h$ is defined as
\begin{equation}\label{topologicalOverlap}
  \omega_{ih}=\frac{l_{ih}+a_{ih}}{\mathrm{min}\{k_i,k_h\}+1-a_{ih}},
\end{equation}
where $a_{ih}=\left(\frac{1+\mathrm{cor}(\bdsl{y}_i, \bdsl{y}_h)}{2}\right)^{\beta}$ is the adjacency measure between genes $i$ and $h$, $l_{ih}=\sum_{u\neq i,h}a_{iu}a_{uh}$ and $k_i=\sum_{u\neq i}a_{iu}$. Moreover, $a_{ih}\in[0,1]$ provides a measure of connectivity between two genes, when they are viewed as two nodes in a network, with $1$ meaning perfectly connected and $0$ meaning not connected. The power parameter $\beta \geq 1$ in $a_{ih}$ provides thresholding by shrinking $a_{ih}$ towards zero, and it needs to be specified a priori. If we use $a_{ih}$ to measure how well genes $i$ and $h$ are connected directly in a network of all the genes, then $\omega_{ih}$ measures how well genes $i$ and $h$ are connected both directly and indirectly though common neighbors in the network. Furthermore, $\omega_{ih}$ takes values in the interval $[0, 1]$, with $1$ meaning that genes $i$ and $j$ are perfectly connected to each other both directly and indirectly through their common neighbors in the network and $0$ meaning they are not connected and do not share any common neighbors. Then, the topological overlap based distance is defined as $ d_{ih}^{\omega}=1-\omega_{ih}$. This procedure is implemented in the \verb'R'  package WGCNA \cite{LangfelderHorvath2008}.

\vskip.2in
\noindent \underline{Dynamic time warping distance}: \\
The dynamic time warping distance is derived under a hierarchical model of curve regisration \cite{TelescaEtal2009}. In the model,  gene $i$ at time $t_j$ can be modeled as,
$ \;   y_{ij}=c_i+a_i m\{\mu_i(t_j)\}+\epsilon_{ij}$,
where $\epsilon_{ij} \iid N(0, \sigma^2)$, $a_i$ and $c_i$ are the amplitude and mean level parameters for the profile of gene $i$, $m()$ is the common shape function shared by all genes and $\mu_i()$ denotes the gene specific time warping function, both of which are modeled with B-splines. Within this framework, the dynamic time warping distance between two genes $i$ and $h$ is defined as,
\begin{equation}\label{timeWarpingDistance}
    D_{ih}=\sum_{j=1}^{m}|\mu_i(t_j)-\mu_h(t_j)|/(t_m-t_1).
\end{equation}
The dynamic time warping distance can be interpreted as the average distance between the timings of the features characterizing the two gene curves. To cluster genes, one could apply hierarchical clustering with the dynamic time warping distance.
In our implementation, we estimate $D_{ih}$ from the raw expression measurements using dynamic programming.

\vskip.2in
\noindent\underline{Autocorrelation based distance}: \\
For gene $i$, the autocorrelation at lag $r \in (1, ..., m-1)$ is given by
\begin{equation}\label{autocorrelation}
    \rho_{ir}=\frac{\sum_{j=r+1}^m(y_{ij}-\bar{y_i})(y_{ij-r}-\bar{y_i})}{\sum_{j=1}^{T}(y_{ij}-\bar{y_i})^2},
\end{equation}
where $\bar{y_i}=\frac{1}{m}\sum_{j=1}^Ty_{ij}$. This technique is based on substituting $\bdsl{\rho}_i=(\rho_{i1}, ..., \rho_{i(m-1)})'$ as the observation vector for gene $i$ \cite{DUrso2009}. $\bdsl{\rho}_i$ could provide additional information about internal associations among observations of gene $i$. WGCNA or other clustering methods are then applied to $\bdsl{\rho}_i$.

\subsection{ Longitudinal representation of cross-sectional data }
In our investigation we consider four case study datasets, acquired cross-sectionally,  with age information available for each subject. For these studies interest centers on the relationship between age and the dynamic evolution of genetic activity. The aim of a clustering exercise is therefore that of identifying subgroups of genes with similar age-average expression or methylation dynamics.

A typical data set is structured as follows. For each gene $i$, $(i=1,.\ldots,p)$, molecular interrogation tissue is obtained from $n$ subjects. We therefore observe an $n-$dimensional expression vector $\tilde{\bf y}_i:n\times 1$, together with age measurements ${\bf x}:n\times 1$ for each subject. We find it useful to represent the data into a pseudo-longitudinal form  with respect to age. We implemented two pre-processing methods: a binning method and a smoothing method. In the binning method, we form age bins $(t_0<t_1<t_2<\ldots<t_m)$ of  fixed bin length. Then, subjects are divided into groups $A_j = \{\ell : x_\ell \in [t_{t-1},t_j) \}$, $(j=1,\ldots,m)$. Within each age bin, for each gene, we define $y_{it_j} = \frac{1}{|A_j|}\sum_{\ell \in A_j } \tilde{y}_{i\ell}$; the mean methylation or expression among subject in the same age bin. In the smoothing method, for each gene, a methylation or expression curve over the age range is estimated by fitting a smoothing spline regression model, s.t. $\tilde{\bf y}_i = f_i({\bf x}) + \epsilon_i(x)$. Smoothing parameters are chosen in a data adaptive manner. Consequently, a pseudo-longitudinal gene profile is defined using predicted values as $y_{it_j} = \hat{f}(t_j)$, computed over all
unique ${\bf x}$ values.

\section{Simulation studies}
\label{sec:simulation}
\noindent \underline{Simulation study 1}\\
The first simulation scheme is adapted from \cite{TelescaEtal2009}. The time course trajectory of gene $i$ in cluster $k$ is generated from $y_{ij}=a_if_k(t_j+\delta_{ij})+\epsilon_{ij}$, where $\epsilon_{ij} \iid N(0, 0.4^2)$, $\delta_{ij} \iid U[-1,1]$ and $a_i \iid N(1, 0.2^2)I(a_i>0)$. The sampling time points, $t_j$, are integers within the interval $T=[0, 30]$. To test the performance of the methods on clustering gene curves with different lengths, we simulate data at every 1, 2, 3 and 6 time points for simulated gene curves. Namely, we use sampling intervals 1, 2, 3 and 6, each of which corresponds to a simulation scenario. For each simulation scenario, we generate 100 datasets. For each simulated dataset, we use $5$ clusters, and the mean curve functions of the first $4$ clusters take on one of the following 4 functions respectively,
$$\begin{array}{ll}
   f_1(t) = \sin((t+0.5)/4))+\cos((t-1)/5, &
   f_2(t) = \cos(t/4),\\
   f_3(t) = -f_1(t), &
   f_4(t) = - f_2(t).
\end{array}$$
The fifth cluster consists of noise genes which are generated from $\mathrm{N}_m(c_i\bdsl{1}, 0.4^2\bdsl{I})$, where $\bdsl{1}$ is a vector of $1$'s, $\bdsl{I}$ is an identify matrix and $c_i \iid U(-1, 1)$. For the five clusters, we generate 50, 100, 200, 300 and 400 gene curves in each dataset in the simulation scenarios with sampling intervals 3 and 6, and generate 25, 50, 100, 150 and 200 gene curves in each dataset in the scenarios with sampling intervals 1 and 2. For scenarios with sampling intervals 1 and 2, cluster sample sizes, 25, 50, 100, 150 and 200, are used due to the time limit allowed by the cluster server used for the simulation study. Simulated gene curves in one simulated dataset are shown in the  Supplementary File. Comparing the cluster mean curves, we notice that the shapes of gene curves from the five clusters do not resemble each other.

To compare the performances of the methods, we use the adjusted Rand index \cite{HubertArabie1985} to measure agreement between the clustering results and true cluster membership. The adjusted Rand index takes on values within interval $[0,1]$, with 1 indicating perfect agreement between two clusterings and 0 indicating two independent clusterings. Details about adjusted Rand index are described in the Supplementary file. In our comparison, we define performance of a method as the median of its adjusted Rand indices and we define stability as the interquartile range of its adjusted Rand indices.

\vskip.2in
\noindent \underline{Simulation study 2}\\
The second simulation scheme is adapted from \cite{InoueEtal2007}. The time course trajectory of gene $i$ in cluster $k$ is generated as $\bdsl{y}_i \sim \mathrm{N}(\bdsl{\theta}_{[k]}', 0.4^2\bdsl{I})$, where $\bdsl{\theta}_{[k]}$ is the $k$th row of matrix $\Theta=[\bdsl{\theta}_0,\ldots,\bdsl{\theta}_j,\ldots,\bdsl{\theta}_{20}]$ and $\bdsl{\theta}_0=(1, 0, 0, 0)'$, $\bdsl{\theta}_j=\bdsl{G}\bdsl{\theta}_{j-1}$ with $\bdsl{G}$ being a $4\times4$ transition matrix defined below. Thus, the sampling time points, $t_j$, are integers with the interval $T=[0, 20]$. To test the performance of the methods on clustering gene curves with different lengths, we simulate data at every 1, 2 and 4 time points for simulated gene curves. Thus, we use sampling intervals 1, 2 and 4, each of which corresponds to a simulation scenario. For the three scenarios, the matrix $\Theta$ is defined as $[\bdsl{\theta}_0, \bdsl{\theta}_1, \ldots, \bdsl{\theta}_{19}, \bdsl{\theta}_{20}]$, $[\bdsl{\theta}_0, \bdsl{\theta}_2, \ldots, \bdsl{\theta}_{18}, \bdsl{\theta}_{20}]$ and $[\bdsl{\theta}_0, \bdsl{\theta}_4, \ldots, \bdsl{\theta}_{16}, \bdsl{\theta}_{20}]$ respectively. For each simulation scenario, we generate 100 datasets. For each simulated dataset, we use 5 clusters, and the mean curve vectors of the first 4 clusters are the row vectors, $\bdsl{\theta}_{[1]}$, $\bdsl{\theta}_{[2]}$, $\bdsl{\theta}_{[3]}$ and $\bdsl{\theta}_{[4]}$, of matrix $\Theta$. We define transition matrix $\bdsl{G}$ as
\begin{equation*}
\bdsl{G}=\left[\begin{array}{cccc}
0.8 & 0.8 & -0.8 & -0.6 \\
0.0 & 0.6 & 0.4 & 0.0 \\
-0.1 & 0.0 & 0.8 & 0.4 \\
0.0 & 0.0 & 0.0 & 0.2
\end{array}\right].
\end{equation*}
The fifth cluster consists of noise genes generated as $\bdsl{y_i}\ind \mathrm{N}_m(c_i\bdsl{1}, 0.4^2\bdsl{I})$, where $c_i\iid U[-0.5, 0.5]$. For the five clusters, we generate 50, 100, 200, 300 and 400 gene curves in each dataset in the simulation scenarios with sampling intervals 2 and 4, and generate 25, 50, 100, 150 and 200 gene curves in each dataset in the scenario with sampling interval 1. For scenario with sampling interval 1, cluster sample sizes, 25, 50, 100, 150 and 200, are used due to the time limit allowed by the cluster server used for the simulation study. Simulated gene curves in one simulated dataset are shown in the Supplementary File. Comparing the cluster mean curves, we notice that the shapes of gene curves from the 5 clusters do not resemble each other except for 2 clusters colored by green and pink.

\begin{table}
\begin{center}
\begin{footnotesize}
\caption{\label{simulTable}Adjusted Rand Indices and computing time for methods and simulations}
\begin{tabular}{|r|*{6}{c}|}
\hline & SSC & FCM & MCLUST & WGCNA$_1$ & DTW & WGCNA$_2$\\ 
 \hline\hline
     ARI & 0.69 & 0.73 & 0.73 & 0.72 & 0.65 & 0.26 \\
S1I6 IQR & (0.67, 0.70) & (0.72, 0.74) & (0.64, 0.74) & (0.70, 0.74) & (0.62, 0.68) & (0.25, 0.26) \\
    time (min) & 291 & 239 & 0.228 & 0.0793 & 18.5 & 0.0985 \\\hline
     ARI & 0.73 & 0.78 & 0.78 & 0.74 & 0.81 & 0.42 \\
S1I3 IQR & (0.71, 0.78) & (0.77, 0.78) & (0.77, 0.78) & (0.72, 0.75) & (0.78, 0.82) & (0.41, 0.43) \\
    time & 548 & 817 & 0.345 & 0.292 & 42.2 & 0.126\\\hline
     ARI & 0.74 & 0.79 & 0.78 & 0.93 & 0.90 & 0.46\\
S1I2 IQR & (0.72, 0.91) & (0.78, 0.79) & (0.78, 0.79) & (0.91, 0.94) & (0.87, 0.92) & (0.45, 0.47) \\
    time & 70.0 & 127 & 0.140 & 0.027 & 14.2 & 0.027 \\\hline
     ARI & 0.74 & 0.79 & 0.80 & 0.99 & 0.91 & 0.49 \\
S1I1 IQR & (0.71, 0.89) & (0.79, 0.80) & (0.79, 0.80) & (0.98, 0.99) & (0.88, 0.94) & (0.48, 0.50) \\
    time & 211 & 378 & 0.165 & 0.0369 & 32.5 & 0.0368 \\\hline
     ARI & 0.65 & 0.66 & 0.64 & 0.62 & 0.55 & 0.15 \\
S2I4 IQR & (0.62, 0.69) & (0.64, 0.68) & (0.61, 0.65) & (0.61, 0.63) & (0.49, 0.60) & (0.14, 0.16) \\
    time & 298 & 263 & 0.422 & 0.0769 & 18.5 & 0.0970 \\\hline
     ARI & 0.74 & 0.67 & 0.67 & 0.72 & 0.77 & 0.45\\
S2I2 IQR & (0.70, 0.78) & (0.66, 0.74) & (0.67, 0.74) & (0.72, 0.73) & (0.75, 0.79) & (0.44, 0.46) \\
    time & 583 & 720 & 0.637 & 0.0803 & 39.9 & 0.0875 \\\hline
     ARI & 0.83 & 0.70 & 0.70 & 0.75 & 0.72 &  0.48\\
S2I1 IQR & (0.75, 0.92) & (0.69, 0.71) & (0.69, 0.71) & (0.74, 0.75) & (0.69, 0.75) & (0.46, 0.48) \\
    time & 187 & 221 & 0.276 & 0.0299 & 18.8 & 0.0283\\\hline
\end{tabular}
\end{footnotesize}
\end{center}
\smallskip
S1I6, S1I3, S1I2 and S1I1: simulation scenarios with sampling intervals 6, 3, 2 and 1 in simulation study 1; S2I4, S2I2 and S2I1: simulation scenarios with sampling intervals 4, 2 and 1 in simulation study 2; ARI: median of adjusted Rand indices; IRQ: interquartile range of adjusted Rand indices; MFDA: model based functional data analysis; FCM: functional clustering model; MCLUST: model-based clustering; DTW: dynamic time warping distance based clustering; WGCNA: weighted gene co-expression network analysis; ACF: autocorrelation based clustering; median and interquartile range of adjusted Rand indices and computation time in minutes are reported for each method for simulation studies 1 and 2 with various sampling intervals.

\end{table}

\begin{figure}
\centerline{\includegraphics[width=165mm]{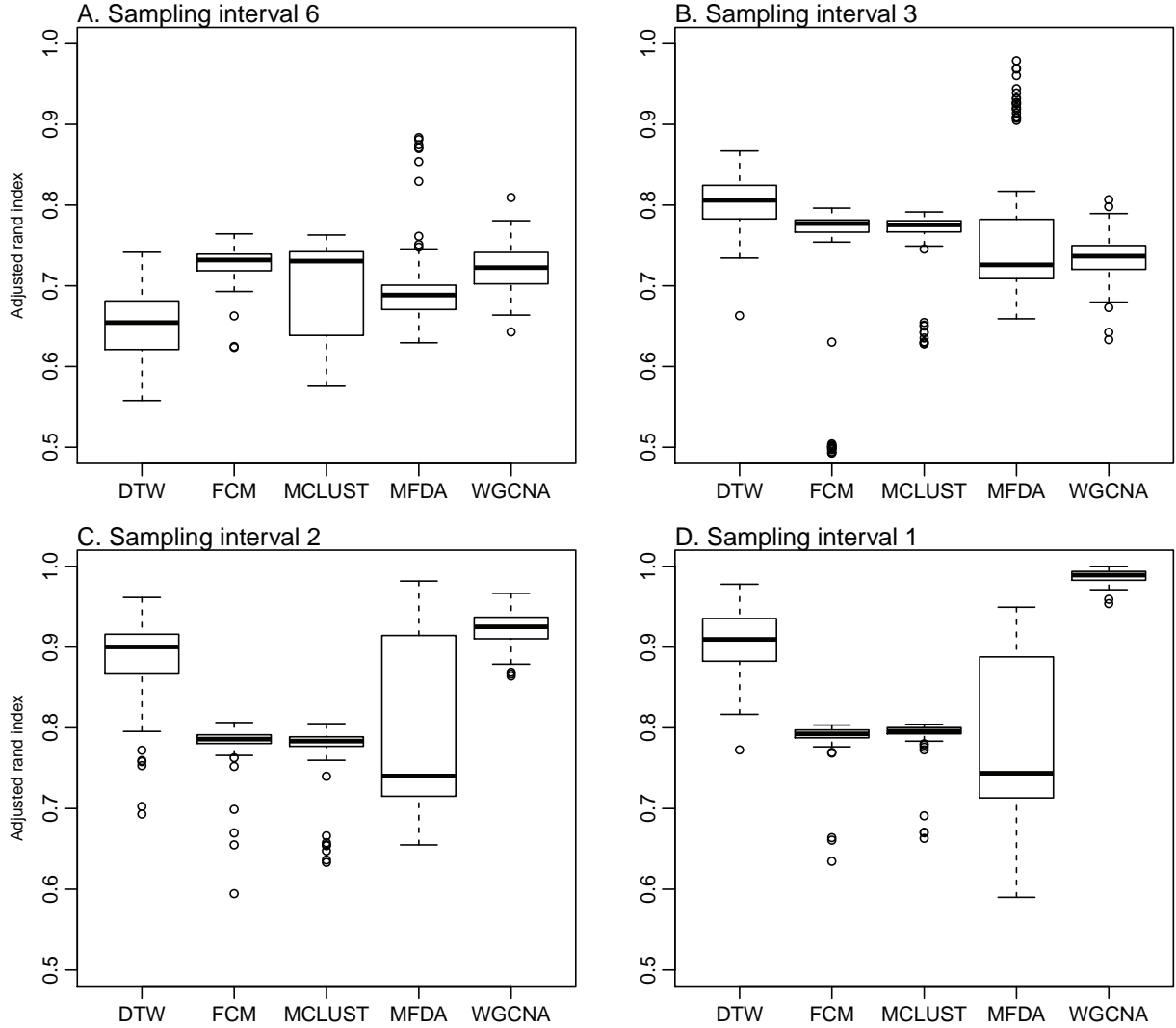}}
\caption{\label{simul1Plot}\textbf{Boxplot of adjusted Rand indices for simulation study 1} Boxplots of adjusted Rand indices of DTW, FCM, MCLUST, MFDA and WGCNA on 100 simulated datasets. (A) Simulation scenario with sampling interval 6. (B) Simulation scenario with sampling interval 3. (C) Simulation scenario with sampling interval 2. (D) Simulation scenario with sampling interval 1. When gene curves are short, FCM gives the best performance. When gene curves are medium, DTW gives the best performance. When gene curves are long ($\geq$ 15 observations), WGCNA gives the best performance. The performance of FCM is very stable with narrow interquartile range of adjusted Rand indices.}
\end{figure}

\begin{figure}
\centerline{\includegraphics[width=165mm]{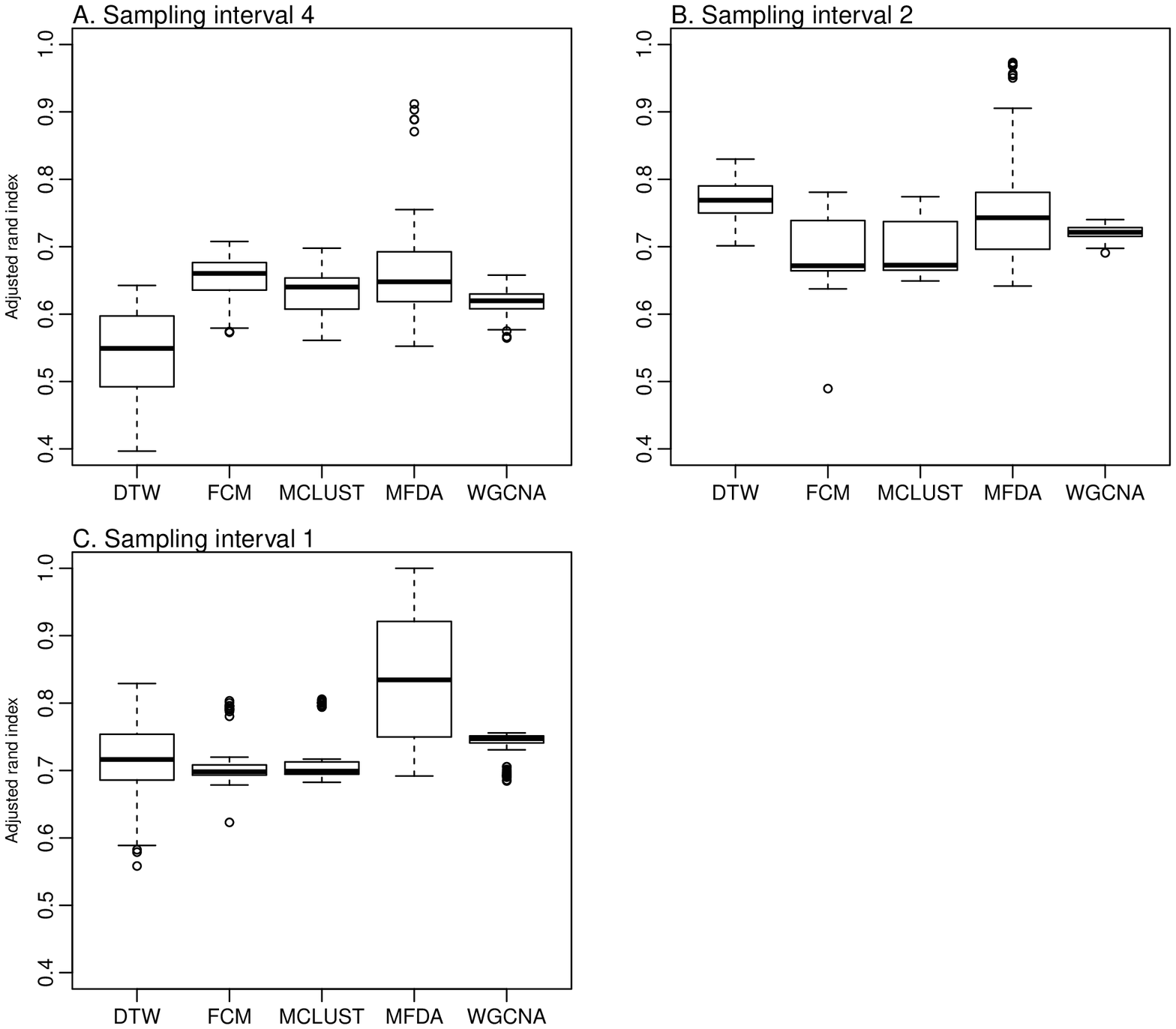}}
\caption{\label{simul2Plot}\textbf{Boxplot of adjusted Rand indices for simulation study 2} Boxplots of adjusted Rand indices of DTW, FCM, MCLUST, MFDA and WGCNA on 100 simulated datasets. (A) Simulation scenario with sampling interval 4. (B) Simulation scenario with sampling interval 2. (C) Simulation scenario with sampling interval 1. MFDA gives good performance for gene curves with all lengths. FCM gives the best performance for short gene curves. DTW gives the best performance for gene curves medium in length. }
\end{figure}

\vskip.2in
\noindent \underline{Simulations results}\\
We apply the six methods discussed in section \ref{clustMethod} on the 100 datasets in each simulation scenario in simulation study 1 and simulation study 2, and compute adjusted Rand indices between the clustering results and the true clustering defined in the simulation studies.  We list the median and inter quartile range of adjusted Rand Indices in Table \ref{simulTable}, tabulated by the six methods and simulation scenarios. We also create boxplots of adjusted Rand indices generated by applying each clustering method on each simulation scenario. The boxplots of adjusted Rand indices from the first simulation study are shown in Figure \ref{simul1Plot} and those from the second simulation study are shown in Figure \ref{simul2Plot}. We summarize the results as follows: when gene curves are short (around 5 observations), FCM gives the best performance, since the model is specifically designed for sparsely sampled data; when curves are medium in length (around 10 observations), DTW gives the best performance. When curves are fairly long ($\geq$ 15 observations), WGCNA gives the best performance. When gene curves are long, formula (\ref{topologicalOverlap}) leads to good estimates of topological overlap measure between genes. As a result, the clustering results generated by WGCNA are similar to the true clustering defined in the simulation studies. For data with gene clusters whose profiles are similar, MFDA performs very well, if not the best, regardless of the length of curves. Results generated by ACF are much worse than the other methods, so we do not include it in the boxplots.

In terms of computing time (Table \ref{simulTable}), both WGCNA and MCLUST are very fast, and they take less than 1 minute to cluster 1050 genes, 11 observations. DTW is slower and takes about 20 to 40 minutes to analyze a dataset. FCM and MFDA are very slow (200 min.). Computation is carried out on a cluster server.

\section{Case Studies}\label{caseChap2}

\subsection{Datasets}

Table \ref{dataTable} describes the human DNA methylation dataset and 3 human RNA expression datasets that are used in the case studies. The table reports the sample size, microarray technology used to generate the data, citation and public availability of the datasets in databases NCBI-GEO and EMBL-EBI.

\begin{table}
\caption{\label{dataTable}Description of methylation and expression data sets}
\begin{center}
\begin{footnotesize}
\begin{tabular}{|*{5}{l}|}
\hline Dataset & Sample size & Platform & Reference & Public availability\\
\hline\hline
Dutch methylation & 658 & Illumina Infin 27 & \cite{HorvathEtal2012} \cite{HorvathEtal2012} & GSE41037 \\
Dutch expression  & 151 & Illumina H-8/H-12 &  \cite{VanEijk2012} & N/A \\
DILGOM expression & 518 & Illumina H-12 &  \cite{Inouye2010} & E-TABM-1036\\
SAFHS expression & 1038 & Illumina WG-6 &  \cite{Goring2007} & E-TABM-305 \\\hline
\end{tabular}
\end{footnotesize}
\end{center}
\smallskip
The table describes four dataset used in the case studies, in terms of the sample size, microarray technology used to generate the data, citation and public availability in databases NCBI-GEO and EMBL-EBI
\end{table}

Dataset 1 has DNA methylation data in whole blood containing polymorphonuclear leukocytes, mononuclear cells, platelets and red
blood cells. DNA methylation microarray data measures methylation levels of genes. A gene is methylated when a methyl group is attached to its DNA sequence, which, in turn, stops the gene's expression and its biological function. Platelets and red blood cells do not contain nuclear DNA. Samples were collected between 1 January 2004 and 31 December 2007 at the University Medical Center Utrecht. Dataset 1 consists of 92 healthy Dutch subjects who had been collected as healthy controls for a case control study of amyotrophic lateral sclerosis and 273 healthy controls and 293 diseased individuals from a case-control study of schizophrenia. Age was also collected on every subject when they entered the study, and the age range is from 16 to 88 with a median of 33. The dataset was collected to study the effects of human aging process and schizophrenia on DNA methylation. By comparing clustering analysis results obtained from this dataset and other datasets collected on human blood and brain samples, the authors found a cluster of genes whose methylation levels increase as age increases consistently across different human tissues, and they also found that correlations between age and DNA methylation are very similar between healthy controls and diseased individuals \cite{HorvathEtal2012}. Here, we are also interested in studying the relation between age and DNA methylation data. Methylation data was generated using the Illumina Infinium Human Methylation27 BeadChip (Illumina Inc., San Diego, CA, USA).

Dataset 2 contains RNA expression in whole blood for a subset of the subjects in dataset 1, specifically, 76 healthy controls from the study of amyotrophic lateral sclerosis and 75 healthy controls from the study of schizophrenia. The median age for subjects in this dataset is 55 and age range is from 19 to 88. The dataset was collected to study the relation between DNA methylation and RNA expression, and it was found that genetic variation, namely single nucleotide change in DNA sequence, affects DNA methylation which further affects RNA expression \cite{VanEijk2012}. In our study, we are interested in studying the relation between RNA expression and age. Gene expression data was generated using Illumina HumanHT-8 beadchip and Illumina HumanHT-12 beadchip (Illumina Inc., San Diego, CA, USA).

Dataset 3 contains RNA expression in whole blood samples from a Finnish cohort of 518 individuals, 240 males and 278 females and within age range from 30 to 70 and median age 50 \cite{Inouye2010}. The cohort was from the Dietary, Lifestyle, and Genetic determinants of Obesity and Metabolic syndrome (DILGOM) study. Originally, the data was collected to study the relation RNA expression and blood lipids. Here, we are interested in studying the relation between RNA expression and age. The RNA expression data was generated using Illumina HumanHT-12 Expression BeadChips (Illumina Inc., San Diego, CA, USA).

Dataset 4 contains RNA expression in peripheral blood samples from 1038 individuals in the San Antonio Family Heart Study (SAFHS), a study that investigated the genetics of cardiovascular disease in Mexican Americans \cite{Goring2007}. In the 1038 individuals, there were 603 females and 435 males, aged from 15 to 94 with median age 37. The data was collected to study genetic aspects of cardiovascular disease, i.e. RNA expression difference between healthy subjects and subjects with cardiovascular disease. We are interested in the relation between age and RNA expression data. RNA expression data was generated using Illumina Sentrix Human Whole Genome (WG-6) Series I BeadChips (Illumina Inc., San Diego, CA, USA).

The four datasets described above are four cross-sectional datasets where gene expression or DNA methylation were measured for 4 cohorts of human subjects, and age of each subject was also collected. To compare the clustering methods for time course microarray data on the four datasets, we apply the binning method and smoothing method discussed in section \ref{dataTransform} and transform the four cross-sectional datasets in to pseudo-longitudinal datasets. If we look at a cross-sectional dataset as a matrix where each column corresponds to a gene's expression levels measured on all subjects and each row corresponds to all genes' expression levels of a subject, then the transformed pseudo-longitudinal dataset is a matrix where each column corresponds to a gene's expression levels measured over different age points and each row corresponds to all genes' expression levels at a specific age point. Because there are overall 20,000 genes in each dataset, computation time for FCM, MFDA and DTW would be several days, which makes them impractical to use. Therefore, we only compare MCLUST and WGCNA in two case studies by applying them to the datasets and testing their ability to identify clusters related to aging, i.e., clusters in which genes' expression or methylation data is highly correlated with age.

\subsection{Comparing Gene Clusters to Polycomb Group Protein Targets}

In the first study, we apply MCLUST and WGCNA on the Dutch methylation dataset to test their ability to identify genes clusters related to aging and compare them with genes that are Polycomb group protein targets (PCGTs). PCGTs is a group of genes which are known to be related to aging process, because they are far more likely to become methylated when a person gets older than non-targets \cite{Teschendorff2010}. Genes in aging related clusters identified by MCLUST or WGCNA are compared to PCGT genes to see how many genes in the aging related clusters overlap with PCGT genes. To evaluate this comparison, we use four statistics, sensitivity, specificity, accuracy and precision, which are defined as follows. Sensitivity is the proportion of genes in both the aging related cluster and PCGTs over genes in PCGTs. Specificity is the proportion of genes not in the aging related cluster and not in PCGTs over genes not in PCGTs. Accuracy is the proportion of genes in the aging related cluster and PCGTs and genes not in the aging related cluster and not in PCGTs over all genes. Precision is the proportion of genes in the aging related cluster and PCGTs over genes in the aging related cluster. For a cluster identified by WGCNA, an eigengene is defined as the first principle component of the matrix with data for genes in the cluster, and it can be considered as the representative pseudo gene for the cluster. For a cluster identified by MCLUST, the estimated mean profile vector of the multivariate normal distribution that is used to model genes in this cluster (see formula \ref{mbcModel}) is defined as the representative pseudo gene for the cluster. To ensure a fair comparison, for WGCNA, we use the cluster whose eigengene has the largest correlation in absolute value with age as the aging related cluster; for MCLUST, we use the cluster whose mean profile vector has the largest correlation in absolute value with age as the aging related cluster. When using the binning method for data transformation, we use 6 bin lengths of 2, 3, 5, 7, 10 and 15 years to create 6 pseudo longitudinal datasets. When using the smoothing method for data transformation, we generate one dataset with DNA methylation levels for all genes at the unique age points in the original dataset.

\begin{figure}
\centerline{\includegraphics[width=165mm]{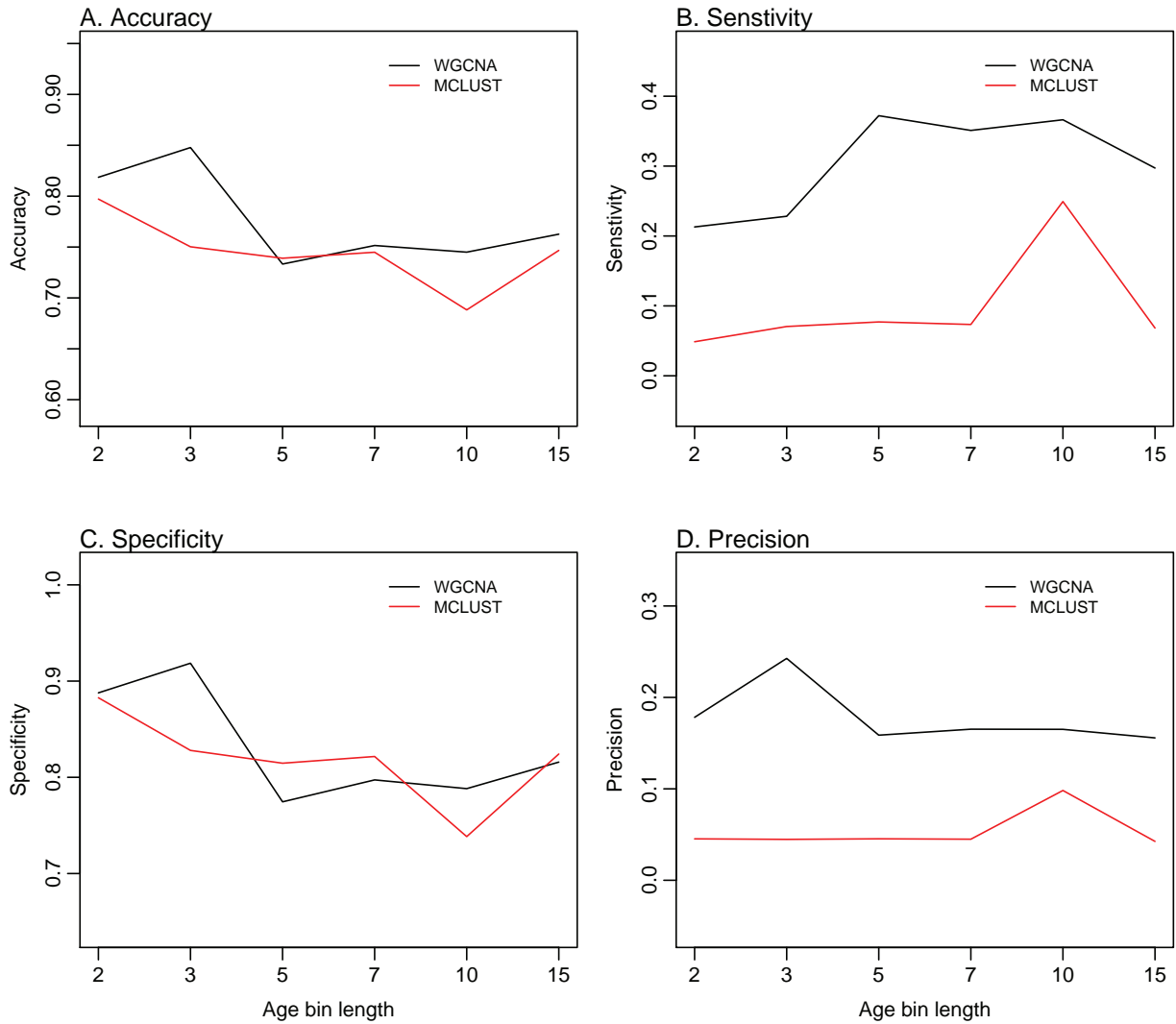}}
\caption{\label{PCGTplot}\textbf{Compare clusters identified by WGCNA and MCLUST to PCGTs} Aging related gene clusters identified by MCLUST and WGCNA are compared to PCGT genes in 6 datasets created with binning methods with bin lengths of 2, 3, 5, 7, 10 and 15 years. Sensitivity, specificity, accuracy and precision are calculated and plotted against bin lengths. WGCNA leads to better sensitivity, precision and accuracy across all the 6 datasets, and ties with MCLUST for specificity.}
\end{figure}

Figure \ref{PCGTplot} shows the results from the datasets generated with the binning method. Sensitivity, specificity, accuracy and precision of comparison between the aging related clusters and PCGTs are calculated in each of the six datasets and then plotted against bin lengths. The larger the bin length is, the fewer data points for each gene in the transformed dataset. Figure \ref{PCGTplot} suggests that the cluster identified by WGCNA has a better agreement with PCGTs than that identified by MCLUST. The cluster identified by WGCNA leads to better sensitivity, precision and accuracy across the 6 datasets, and ties with MCLUST for specificity. Table \ref{PCGTtable} shows the results from the dataset generated by the smoothing method, and sensitivity, specificity, precision and accuracy are calculated. It also suggests that WGCNA does slightly better than MCLUST, having higher accuracy, specificity and precision, but lower sensitivity.

\begin{table}
\caption{\label{PCGTtable} Compare clusters identified by WGCNA and MCLUST to PCGTs}
\begin{center}
\begin{tabular}{|*{5}{l}|}
\hline Method & Accuracy & Sensitivity & Specificity & Precision \\
\hline\hline
WGCNA  & 0.756 & 0.335 & 0.804 & 0.164 \\
MCLUST & 0.741 & 0.347 & 0.786 & 0.156 \\\hline
\end{tabular}
\end{center}

\smallskip
Aging related gene clusters identified by MCLUST or WGCNA is compared to PCGTs in the dataset created with the smoothing method. Sensitivity, specificity, accuracy and precision are reported.
\end{table}

\subsection{Cluster Preservation Study Between Expression Datasets}

In the second case study, we first identify the aging related clusters in the Dutch expression dataset using WGCNA and MCLUST, and then compare their preservation in the DILGOM expression dataset and SAFHS expression dataset. Towards that end, we use the cluster preservation statistic, $Z_\mathrm{summary}$, proposed in \cite{Langfelder2011}. $Z_\mathrm{summary}$ measures if genes in a cluster identified from a reference dataset show high connectivity between each other in a test dataset. It is suggested that if $\mathrm{Z_{summary}}>10$ there is strong evidence that the cluster is preserved; if $2<\mathrm{Z_{summary}}<10$ there is weak to moderate evidence of preservation; if $\mathrm{Z_{summary}}<2$, there is no evidence that the cluster is preserved. Details about the cluster preservation statistic are described in the Supplementary file. In our case, an aging related cluster identified in the Dutch expression dataset is more biologically meaningful, if it is more preserved in the DILGOM expression dataset and SAFHS expression dataset. When using the binning method for data transformation, we use a bin length of 5 years. When using the smoothing method for data transformation, we generate DNA methylation levels for all genes at the unique age points in the original dataset.

\begin{figure}
\centerline{\includegraphics[width=165mm]{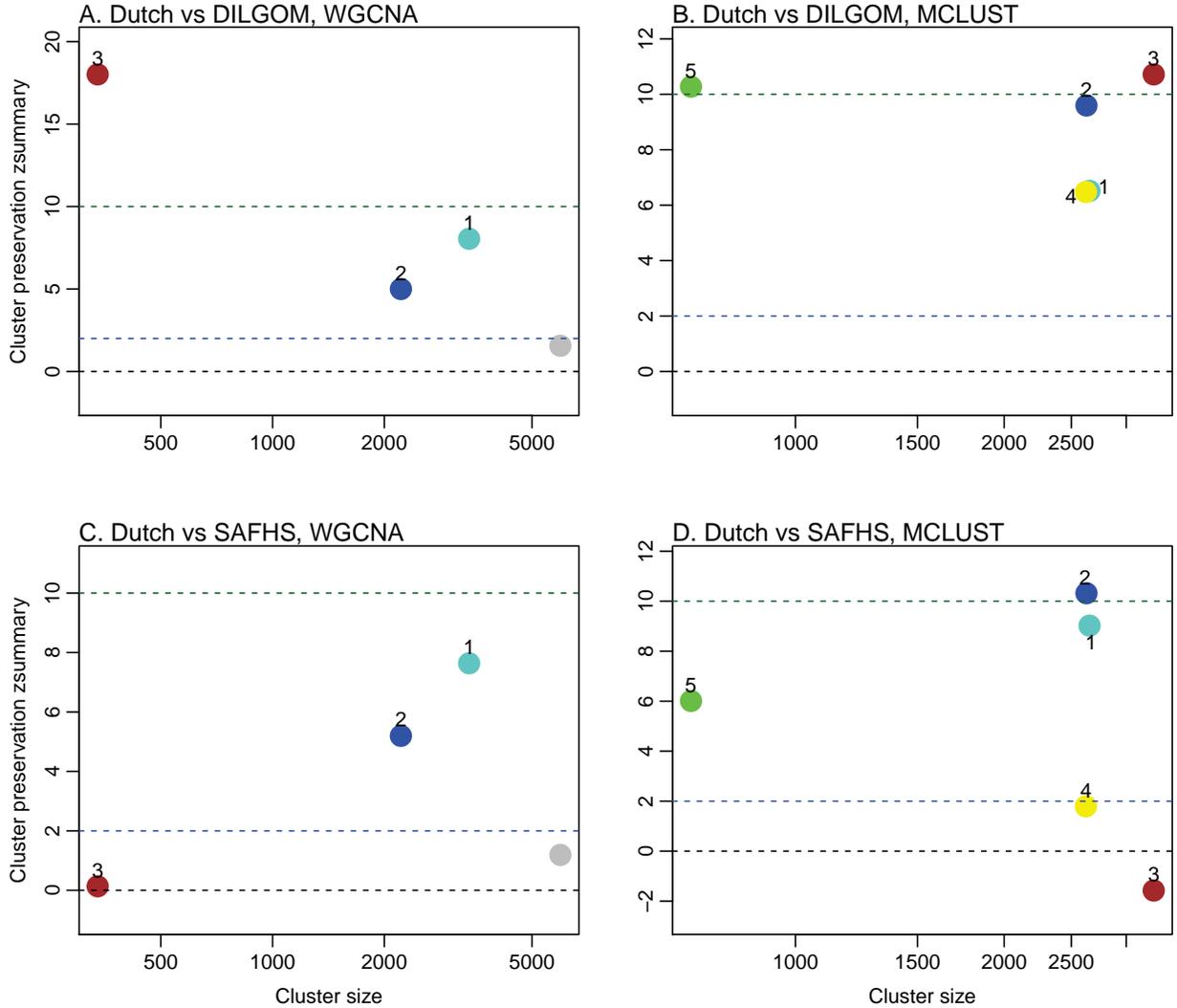}}
\caption{\label{preserv1Plot}\textbf{Cluster preservation on datasets transformed by the binning method} $\mathrm{Z_{summary}}$ are plotted for all clusters.  Aging related clusters are labeled as 3  and colored by brown. (A) Preservation of clusters identified by WGCNA between the Dutch expression dataset and the DILGOM expression dataset. (B) Preservation of clusters identified by MCLUST between the Dutch expression dataset and the DILGOM expression dataset.  (C) Preservation of clusters identified by WGCNA between the Dutch expression dataset and the SAFHS expression dataset. (D) Preservation of clusters identified by MCLUST between the Dutch expression dataset and the SAFHS expression dataset.}
\end{figure}

\begin{figure}
\centerline{\includegraphics[width=16.5cm]{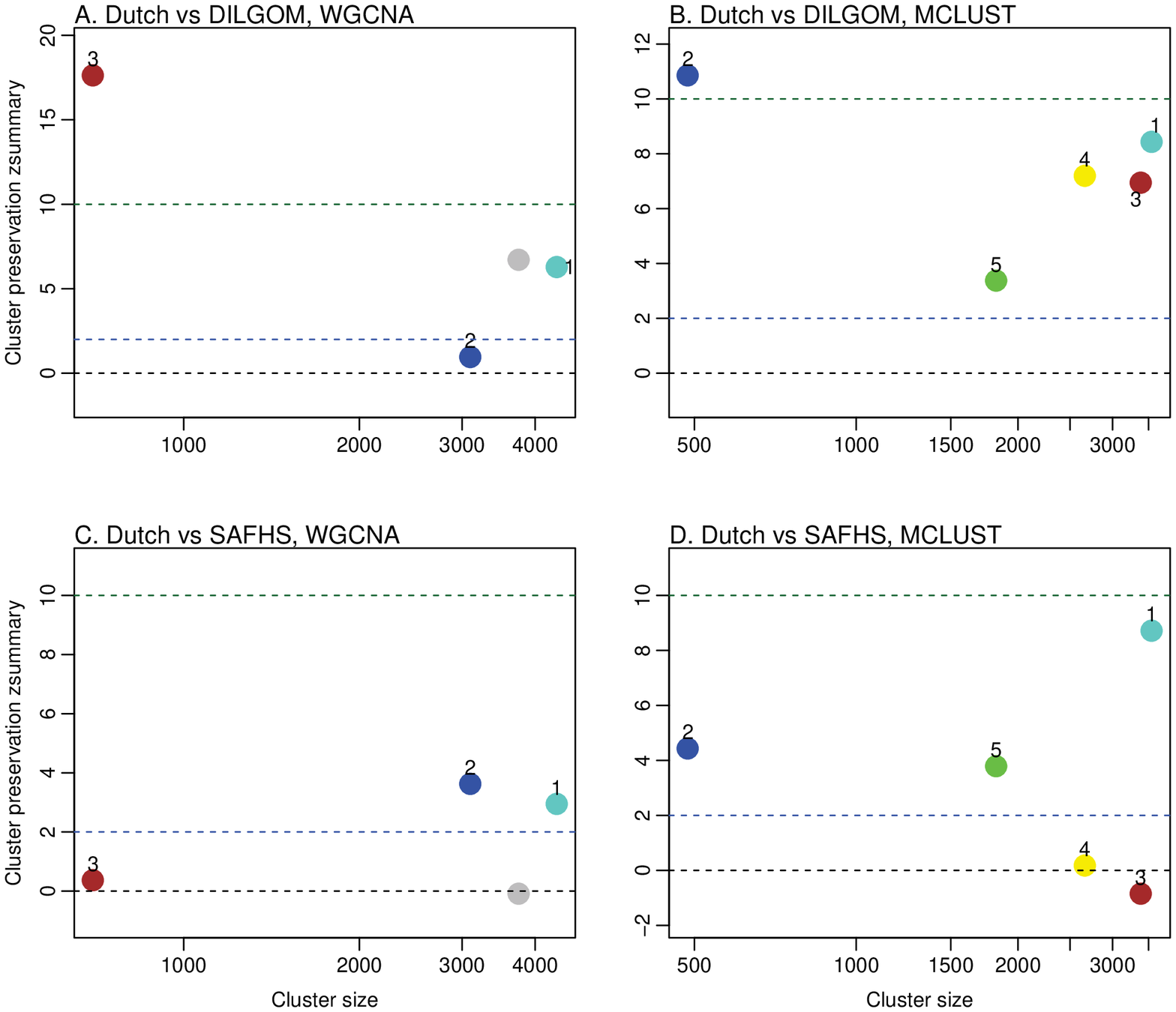}}
\caption{\label{preserv2Plot}\textbf{Cluster preservation on datasets transformed by the smoothing method} $\mathrm{Z_{summary}}$ are plotted for all clusters. Aging related clusters are labeled as 3 colored by brown. (A) Preservation of clusters  identified by WGCNA between the Dutch expression dataset and the DILGOM expression dataset. (B) Preservation of clusters identified by MCLUST between the Dutch expression dataset and the DILGOM expression dataset.  (C) Preservation of clusters identified by WGCNA between the Dutch expression dataset and the SAFHS expression dataset. (D) Preservation of clusters identified by MCLUST between the Dutch expression dataset and the SAFHS expression dataset.}
\end{figure}

Figure \ref{preserv1Plot} shows the cluster preservation results from the datasets generated by the binning method. In the plots, the preservation statistics $\mathrm{Z_{summary}}$ are plotted for all clusters, and the aging related cluster is labeled as 3 and colored by brown. As shown in panel A, cluster 3 identified by WGCNA has $\mathrm{Z_{summary}}$ greater than 10, so it is strongly preserved between the Dutch expression dataset and the DILGOM expression dataset. Similarly, cluster 3 identified by MCLUST is also strongly preserved between the Dutch expression dataset and the DILGOM expression dataset as shown in panel B. As shown in panel C, cluster 3 identified by WGCNA has $\mathrm{Z_{summary}}$ less than 2, so it is not preserved between the Dutch expression dataset and the SAFHS expression dataset. Similarly, cluster 3 identified by MCLUST is not preserved between the Dutch expression dataset and the SAFHS expression dataset as shown in panel D.

Figure \ref{preserv2Plot} shows the cluster preservation results from the dataset generated by the smoothing method. Again, the preservation statistics $\mathrm{Z_{summary}}$ are plotted for all clusters, and the aging related cluster is labeled as 3 and colored by brown. As shown in panel A, cluster 3 identified by WGCNA has $\mathrm{Z_{summary}}$ greater than 10, so it is strongly preserved between the Dutch expression dataset and the DILGOM expression dataset. However, in panel B, cluster 3 identified by MCLUST has $\mathrm{Z_{summary}}$ between 2 and 10, so it is moderately preserved between the Dutch expression dataset and the DILGOM expression dataset. As shown in panel C, cluster 3 identified by WGCNA has $\mathrm{Z_{summary}}$ less than 2, so it is not preserved between the Dutch expression dataset and the SAFHS expression dataset. Similarly, cluster 3 identified by MCLUST is not preserved between the Dutch expression dataset and the SAFHS expression dataset as shown in panel D.

\begin{figure}
\centerline{\includegraphics[width=16.5cm]{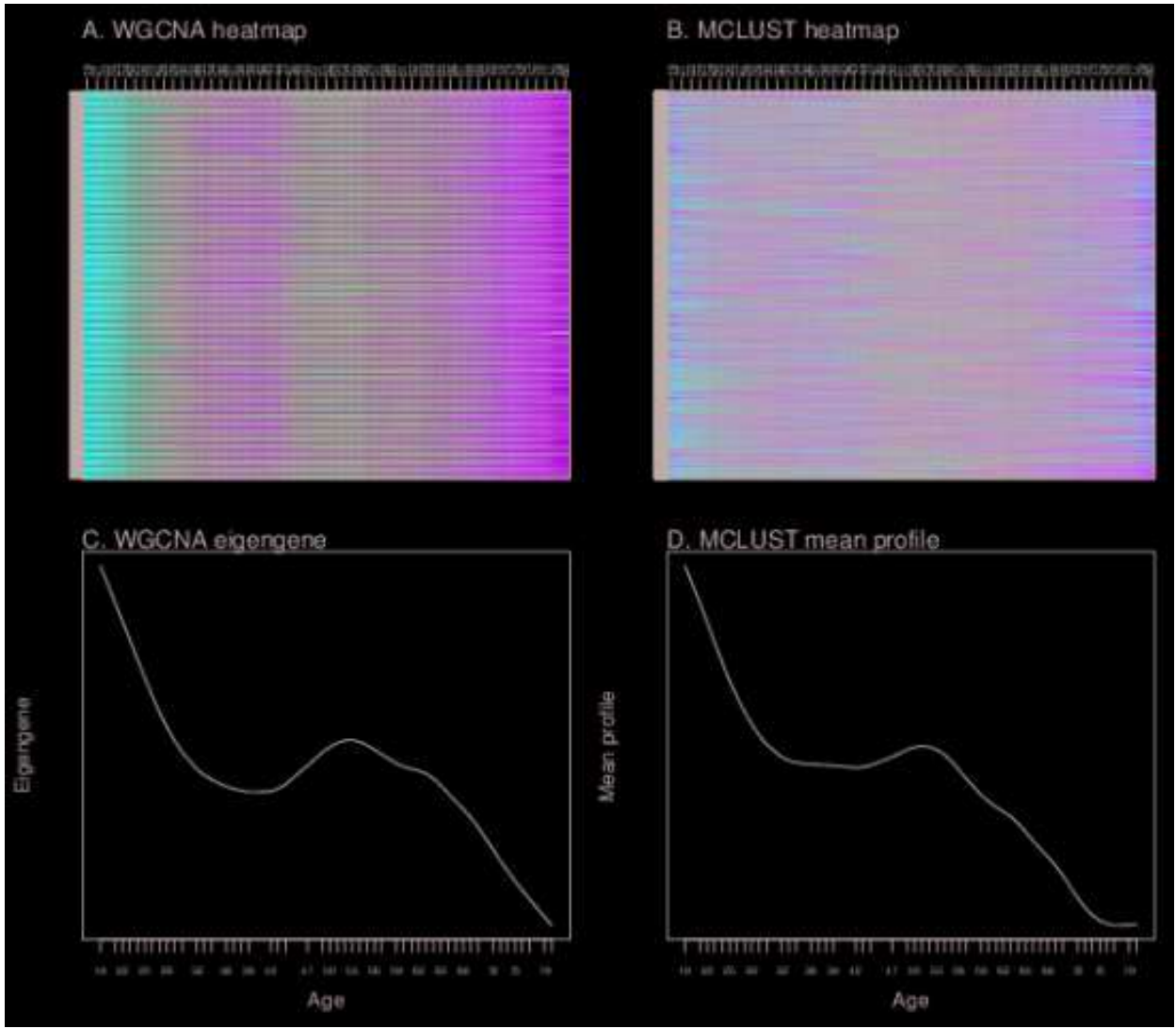}}
\caption{\label{heatmapSmo}\textbf{Heatmap of clusters identified by WGCNA and MCLUST on the Dutch expression dataset transformed by the smoothing method} (A) Heatmap of the aging related cluster identified by WGCNA on the Dutch expression data transformed by the smoothing method, each row represents the expression pattern of a gene with color red meaning high expressions and color green meaning low expressions. Numbers on the top are age points. (B) Heatmap of the aging related cluster identified by MCLUST on the Dutch expression data transformed by the smoothing method. (C) Eigengene of the aging related cluster identified by WGCNA plotted against age. (D) Mean curve of the aging related cluster identified by MCLUST plotted against age. }
\end{figure}

To further study the aging related clusters identified by WGCNA and MCLUST on the Dutch expression dataset transformed by the smoothing method, we plot their heatmaps and the cluster eigengene of the cluster identified by WGCNA and the cluster mean vector of the cluster identified by MCLUST in Figure \ref{heatmapSmo}. For genes in the cluster identified by WGCNA, their expression levels clearly decrease as age increases as shown in panel A. However, such pattern is much weaker for the genes in the cluster identified by MCLUST as shown in the heatmap in panel B. So, the clusters identified WGCNA and MCLUST show very different expression patterns as age increases, although the eigengene of the cluster identified by WGCNA and the mean vector of the cluster identified by MCLUST are very similar. We observe similar results on clusters identified by WGCNA and MCLUST on the Dutch expression dataset transformed by the binning method, as shown in the Supplementary File.

\section{Discussion}\label{diMFDAhap2}
From the simulation study, we find that FCM gives the best performance when gene curves are short, which is consistent with the fact that FCM is specifically designed to handle sparsely sampled functional data with only a few observations for each curve. For gene curves with slightly different shapes, MFDA is able to separate them into different clusters, but the other methods tend to combine those curves into one cluster. DTW gives the best performance when gene curves are medium in length (10 observations). WGCNA performs best when gene curves are long ($\geqslant 15$ observations). MCLUST's performance is very stable.

In terms of computation time, FCM, MFDA and DTW are considerably slower than WGCNA and MCLUST. Computation time depends on the number of genes and length of each gene curve. For example, when there are 1000 genes each of which has 5 observations, MFDA and FCM require 4 to 5 hours of computation time on a node of a cluster server. DTW needs 20 minutes, but WGCNA and MCLUST only need less than 5 seconds. In the case studies, when there are over 20,000 genes in a dataset, FCM, MFDA and DTW require over days of computation time, but WGCNA and MCLUST only need around 100 minutes. The fast computation makes WGCNA and MCLUST a lot more feasible than the other methods. However, functional data model based methods, such as FCM and MFDA, do not require the data to be regularly sampled and deal with the missing data and imputation naturally by using predicted values from the fitted curves. This feature is useful when data are sparsely or irregularly sampled.

By comparing WGCNA and MCLUST on real datasets and testing their ability to identify aging related clusters, we find WGCNA have a slight edge over MCLUST, which is especially supported by the heatmap plot of the aging related cluster identified by the two methods in the Dutch expression dataset. The genes in the cluster identified by WGCNA show very clear association with age, i.e. expression level decreases as age increases. But the genes in the cluster identified by MCLUST show much less clear association with age. However, model based clustering methods, such as MCLUST, are attractive because they provide uncertainty measures of their clustering results and confidence intervals for estimated cluster mean curves.

To transform our cross sectional data into pseudo longitudinal datasets, we implement two methods, the binning methods and the smoothing method. The results from the two methods are similar to each other. We recommend the smoothing method, because it is more convenient to generate a curve over a set of time points using the spline regression model fitted from the original data.

\section{Conclusion}\label{chap2Conclude}
In this chapter, we compare 3 model based clustering methods (MFDA, FCM and MCLUST) with 3 distance based clustering methods (WGCNA, DTW and ACF) for time course microarray data. When both performance and computation time are both taken into consideration, WGCNA and MCLUST perform best, with WGCNA slightly outperforming MCLUST.

\bibliographystyle{plainnat}
\bibliography{ClustComp}

\end{document}